\documentclass[a4paper]{jpconf}
\usepackage{graphicx}
\usepackage{amsmath}
\usepackage{type1cm}

\usepackage{xspace}
\xspaceaddexceptions{]}

\newcommand{\dg}{\ensuremath{^\circ}\xspace}

\newcommand{\al}{\ensuremath{\alpha}\xspace}

\newcommand{\z}{\ensuremath{\zeta}\xspace}

\newcommand{\az}{\ensuremath{(\alpha,\zeta)}\xspace}

\newcommand{\fermi}{\textit{Fermi}~LAT\xspace}

\newcommand{\gr}{\ensuremath{\gamma}-ray\xspace}

\newcommand{\rlc}{\ensuremath{R_{\rm LC}}\xspace}

\newcommand{\figr}[1]{Fig.~\ref{fig:#1}\xspace}
\newcommand{\sect}[1]{Sec.~\ref{sec:#1}\xspace}

\begin{document}
\title{The anatomy of \gr pulsar light curves}

\author{A~S Seyffert$^1$, C Venter$^1$, T~J Johnson$^{2,3}$ and A~K Harding$^4$}
\address{$^1$~Centre for Space Research, North-West University, Potchefstroom Campus, Private Bag X6001, Potchefstroom 2520, South Africa}
\address{$^2$~National Research Council Research Associate}
\address{$^3$~High-Energy Space Environment Branch, Naval Research Laboratory, Washington, DC 20375, USA}
\address{$^4$~Astrophysics Science Division, NASA Goddard Space Flight Center, Greenbelt, MD 20771, USA}
\ead{20126999@nwu.ac.za}

\begin{abstract}
    We previously obtained constraints on the viewing geometries of 6 \fermi pulsars using a multiwavelength approach \cite{Seyffert}. To obtain these constraints we compared the observed radio and \gr light curves (LCs) for those 6 pulsars by eye to LCs predicted by geometric models detailing the location and extent of emission regions in a pulsar magnetosphere. As a precursor to obtaining these constraints, a parameter study was conducted to reinforce our qualitative understanding of how the underlying model parameters effect the LCs produced by the geometric models. Extracting useful trends from the \gr model LCs proved difficult though due to the increased complexity of the geometric models for the \gr emission relative to those for the radio emission. In this paper we explore a second approach to investigating the interplay between the model parameters and the LC atlas. This approach does not attempt to understand how the set of model parameters influences the LC shapes directly, but rather, more fundamentally, investigates how the set of model parameters effects the sky maps from which the latter are extracted. This allows us to also recognise structure within the atlas itself, as we are now able to attribute certain features of the LCs to specific features on the sky map, meaning that we not only understand how the structure of single LCs come about, but also how their structure changes as we move through the geometric solution space.
\end{abstract}

\section{Introduction}

  The \textit{Fermi} Large Area Telescope (LAT) is the primary instrument on the \textit{Fermi Gamma-ray Space Telescope} mission which was launched on 11 June 2008. \fermi is an imaging, wide field-of-view, high-energy \gr telescope covering the energy range from 20~MeV to 300~GeV. The \fermi continuously scans the sky in an ever improving all-sky survey, taking advantage of it's large field-of-view and high sensitivity to scan the entire sky every three hours. One of the important products to date of this all-sky survey is the first \fermi catalogue of \gr pulsars \cite{Abdo} produced using the first six months of \fermi data, and increasing the number of known \gr pulsars from at least 6 \cite{Thompson01} to 46. The second \fermi \gr pulsar catalogue is due to be published soon, and will push the number of known \gr pulsars to over 100. This dramatic increase has opened up a whole new population of pulsars, with exciting possibilities for multiwavelength studies \cite{Weltevrede}.

  We have previously obtained fits to 6 \fermi pulsar LCs using both radio and \gr data to concurrently fit radio and \gr geometric pulsar model LC predictions \cite{Seyffert_b}.

  \subsection{The geometric pulsar models}\label{sec:model}
    We use an idealized picture of the pulsar system, wherein the magnetic field has a retarded dipole structure \cite{Deutsch55} and the \gr emission originates in regions of the magnetosphere (referred to as `gaps') where the local charge density is sufficiently lower than the Goldreich-Julian charge density \cite{GJ69}. These gaps facilitate particle acceleration and radiation. We assume that there are constant-emissivity emission layers embedded within the gaps in the pulsar's corotating frame. The location and geometry of these emission layers determine the shape of the \gr LCs, and there exist multiple models describing their geometry.

    Two such models for the geometry of the \gr emission regions are the outer gap (OG, \cite{CHR86}) and two-pole caustic (TPC, \cite{Dyks03}) models. In both the OG and TPC models emissions are produced by accelerated charged particles moving within narrow gaps along the last open magnetic field lines (the field lines which close at the light cylinder). In the OG model, radiation originates above the null charge surface (where the Goldreich-Julian charge density changes sign) and \textit{interior to} (closer to the magnetic axis) the last open magnetic field lines. The TPC gap starts at the stellar surface and extends \textit{along} the last open field lines up to near the light cylinder, where the corotation speed approaches the speed of light. The special relativistic effects of aberration and time-of-flight delays, which become important in regions far from the stellar surface (especially near the light cylinder), together with the curvature of the magnetic field lines, cause the radiation to form caustics (accumulated emission in narrow phase bands). These caustics are detected as peaks in the observed \gr LCs~\cite{Dyks03,Morini83}.

  \subsection{Phaseplots and LCs}
    The tool we use to visualise the emission of the pulsars is called a phaseplot, and is an equirectangular projection skymap of the emission of the pulsar with the phase at which the emission is observed, $\phi$, on the horizontal axis, and \z on the vertical axis. For each point in the $(\phi,\z)$ space it gives the relative intensity of the emission per solid angle which would be observed if the line-of-sight to the observer would pass through that point. On this phaseplot the pulsar makes one full rotation around its rotation axis, and the equator of the NS is at $\z=90\dg$. To find out what an observer would see during the course of the period we can obtain an LC from this phaseplot by taking a cut at a constant \z through the phaseplot. \figr{labeled} shows an example of a phaseplot (for the TPC case) for a pulsar with an inclination angle of $\al=50\dg$, accompanied by an LC extracted from it using such a cut (designated by the yellow line through the phaseplot). Both the phaseplot and LC are independently normalised so their respective maxima are 1.

\section{Obtaining LC fits\label{sec:obF}}
  To obtain fits to the observed pulsar LCs we need to consider the entire domain of \al and \z at the very least, keeping the other parameters fixed. To most effectively do this we generate `atlases', consisting of a set of model LCs covering the \az domain at a chosen resolution, i.e., a grid of LCs sampling the two-dimensional parameter space evenly. The data are then compared to this atlas (for our purposes this is done by eye) and the best-fit \az is then inferred.

  This fitting process has proven to be effective in a multiwavelength arena with concurrent fitting of both radio and \gr light curves yielding well-constrained best-fit parameters for a number of radio-loud \gr pulsars \cite{Seyffert}. In the case of by-eye fitting the effectiveness and usefulness of this process is quite dependent on the level of the qualitative knowledge available about the behaviour of the produced atlas of LCs when model parameters other than \al and \z are varied. This knowledge can be best garnered by performing parameter studies on the models focusing on the relevant parameters.

  \subsection{Performing parameter studies: radio models vs. \gr models}
    To perform parameter studies using these phaseplots and LCs it is convenient to investigate the interplay between the parameters and the LC atlas. In the case of the simpler conal radio models this approach proves very effective and a lot can be learned from doing such parameter studies. The \gr models yield considerably more complex phaseplots, and as such, more convoluted LC atlases. This means that to obtain useful knowledge from these parameter studies we need to adjust our approach accordingly.

    \begin{figure}[!ht]
      \begin{center}
	\includegraphics[width=0.62\textwidth]{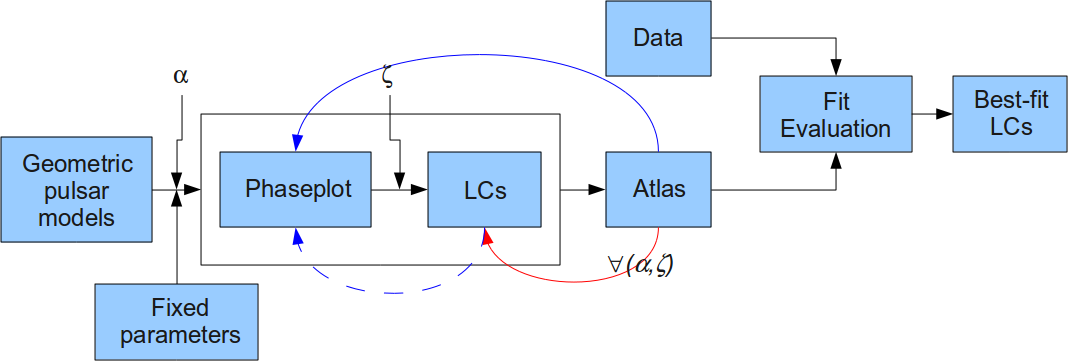}
      \end{center}
      \caption{The process whereby fits to observed LCs can be obtained from implemented geometric models. The red arrow denotes the methedology of understanding the structure of the atlas in light of only the structure of the LCs. The blue lines (solid and dashed combined) denote how we propose this understanding of LC structure could be attained more effectively.\label{fig:flow}}
    \end{figure}

    \figr{flow} shows a schematic representation of the process used to obtain fits to the radio and \gr LCs. The red arrow illustrates the established methodology of understanding the influence the parameters have on the LC atlas, by examining the LCs directly. The blue arrows illustrate the proposed methodology wherein the influence of the model parameters on the LC atlas is understood by examining the effects of the model parameters on the structure of the phaseplots from which the LC atlas is generated.

\section{Re-examining the LCs and phaseplots}
\begin{figure}[!ht]
    \centering
    \begin{minipage}[l]{0.51\textwidth}
      \includegraphics[width=\textwidth]{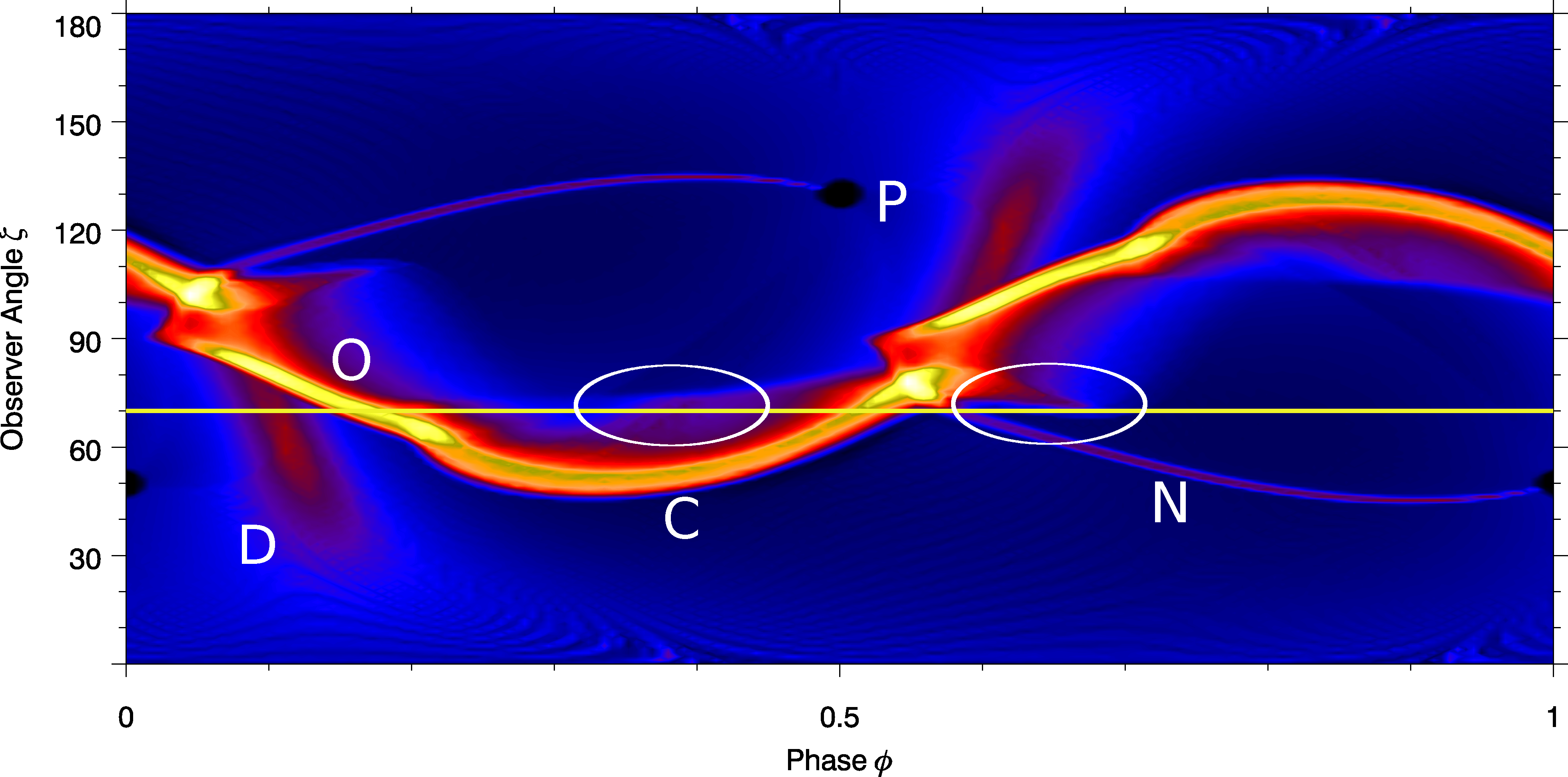}
    \end{minipage}
    \begin{minipage}[r]{0.26\textwidth}
      \includegraphics[width=\textwidth]{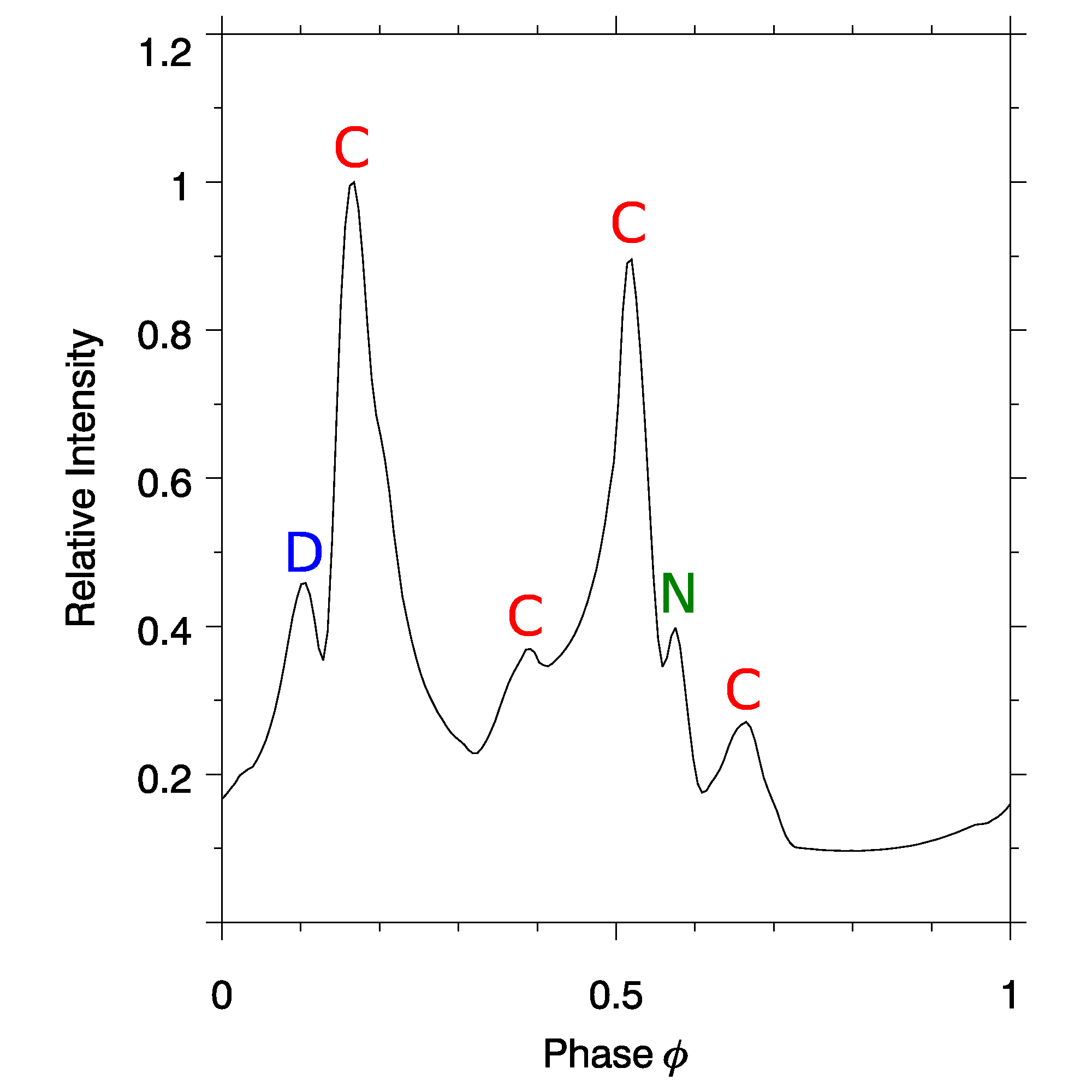}
    \end{minipage}
    \caption{A phaseplot generated by the TPC model with $\al=50\dg$ and cut at $\z=70\dg$. \textit{C} denotes the main caustic, \textit{D} denotes the secondary caustic, \textit{P} denotes the polar cap, and \textit{N} denotes the notch line.\label{fig:labeled}}
  \end{figure}

  \figr{labeled} shows how the different features on the phaseplot can be associated with the peaks seen on the LC. In this case (the TPC model with $\al=50\dg$ and $\z=70\dg$) the two main peaks on the LC are produced by cutting the caustic (marked \textit{C} on the phaseplot) above its lowest point (around $\z=50\dg$), thus intersecting it twice. The earlier (at lower $\phi$) of these two peaks also includes emission from the so-called `overlap region' (where the directions of many of the rotationally distorted magnetic field lines are such that their emission overlaps; marked \textit{O} on the phaseplot). The LC furthermore includes four smaller peaks, one preceding the earliest large peak, one between the two large peaks, and two following the later large peak. The earliest of the four is associated with the so-called `dynamical caustic' (marked \textit{D} on the phaseplot), and is the dominant peak for LCs produced cutting below (at smaller \z) the lowest point of the caustic. The second and fourth peaks of the four smaller peaks are associated with the caustic itself, respectively being produced by very high altitude emission (where the magnetic field is most distorted), and by the small tail of the caustic (both encircled). The third of the four smaller peaks is produced by the line on the phaseplot associated with the so-called `notch' in the rim of the neutron star's polar cap at the surface.

  Thus, we see that the parts of an extracted LC can be understood in terms of the features on the phaseplot which generate them. This means that the hard work of understanding how the LCs change under changes in the underlying parameter set, shifts to having to understand how the structure of the phaseplots change under changes in the underlying parameters. There are mainly 5 features on the phaseplot we need to account for, namely:

  The \textit{main caustic, C}: This feature is the result of bunching of emission directions occurring due to aberration and time-of-flight effects.

  The \textit{secondary (dynamical) caustic, D}: This feature is brought about purely through the effects of aberration and time-of-flight, and at higher \al merges with the above caustics. This feature is accompanied by a spreading of emission directions in the rest of the rotation ($(\phi,\z)$ between $(0.3,50\dg)$ and $(1,0/dg)$; a \textit{trough}).

  The \textit{polar cap, P}: The polar cap is delimited by the footpoints of the last magnetic field lines to close within the light cylinder. Due to the screening of the electric field above the polar cap by the particles produced in the pair cascade, no radiation is expected above the polar cap. In LCs cutting this feature a very sharp fall in emission intensity can be observed at the appropriate phases.

  The \textit{overlap region, O}: This region is where the distortion of the magnetic field under the rotation of the pulsar causes emission directions to overlap. It is associated with the notch in the polar cap rim.

  The \textit{notch line, N}: This line on the phaseplot is a result of the bunching of emission directions due to the bunching of the magnetic field lines of the pulsar.

\section{Revisiting the Atlas}
  \subsection{A single phaseplot}
\begin{figure}[!ht]
      \centering
      \includegraphics[width=0.85\textwidth]{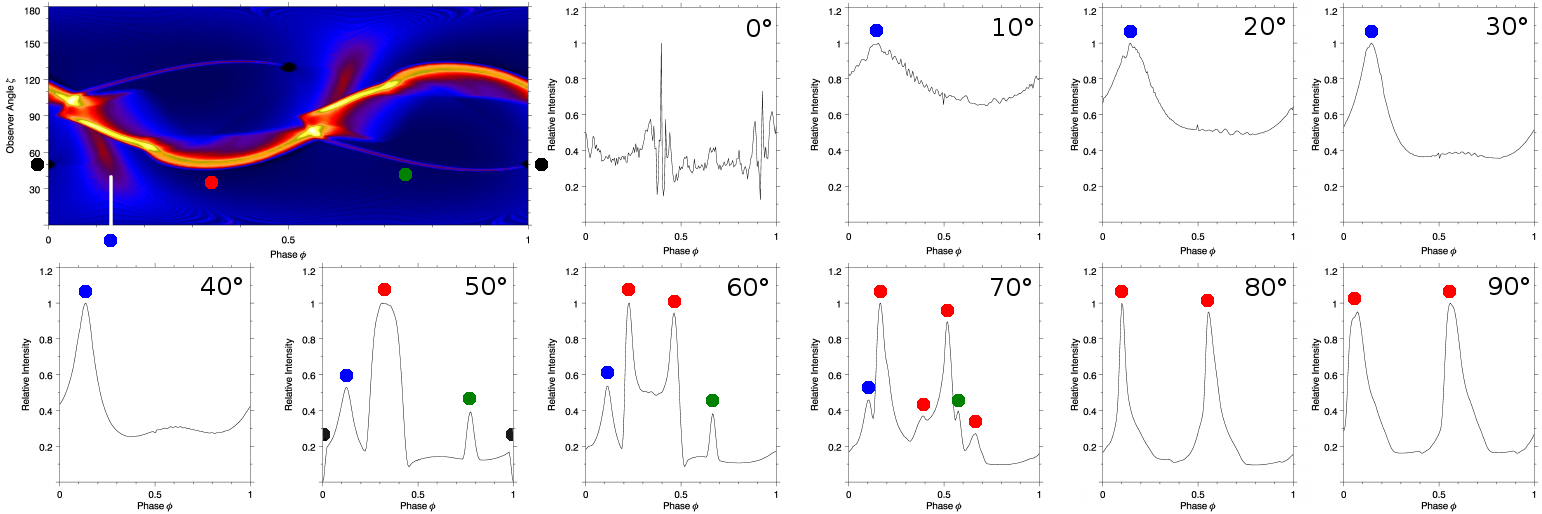}
      \caption{The set of LCs produced by cutting the same phaseplot (as in Fig. \ref{fig:labeled}) at successively higher \z, starting at $\z=0\dg$, and going as high as $\z=90\dg$. Note that the phaseplot is symmetric under a 0.5 shift in phase accompanied by a reflection about the equator due to the symmetry of the pulsar's magnetic field (between hemispheres). This means that cuts above $\z=90\dg$ are redundant.\label{fig:pp_to_lc}}
    \end{figure}

    The LC atlas is produced by generating a phaseplot for each step in the range of possible \al, and for each of those phaseplots cutting it at each step in the range of possible \z (see \sect{obF}). \figr{pp_to_lc} shows the same phaseplot as is used in \figr{labeled}, for multiple \z cuts. The colours of the dots indicate which feature on the phaseplot the LC peaks are associated with.

    We can now very clearly see the structure of the phaseplot reflected in the set of LCs produced from it. We see that at $\z\leq40\dg$ the only dominant peak is associated with the secondary caustic (blue dots), except for $\z=0\dg$, where the LC is entirely made up of low level emission. At $\z=50\dg$ we see that (besides coincidentally cutting through the polar cap (black dots) and seeing the accompanying dip in intensity at $\phi=0$ and 1) the main caustic (red dots) suddenly becomes the dominant feature on the LC as we cut it tangentially at its lowest point. Furthermore, we also see the notch line (green dots) for the first time. From $\z=60\dg$ onward we see that predominantly two-peaked profiles are produced as we are now cutting the caustic above its lowest point and our line-of-sight is intersecting with it twice. At $\z=80\dg$ and 90\dg we see pure two-peaked profiles. Notice the movement of the peak associated with the notch line moving toward earlier phases as we increase \z until its effect becomes lost under the intense emission coming from the caustic. Also, notice that the separation between the two dominant peaks ($\bigtriangleup$) steadily increases to almost 0.5 in $\phi$ as we progressively cut higher above the lowest point of the caustic.

  \subsection{The entire set of phaseplots}
\begin{figure}[!ht]
    \centering
    \begin{minipage}[l]{0.32\textwidth}
      \includegraphics[width=\textwidth]{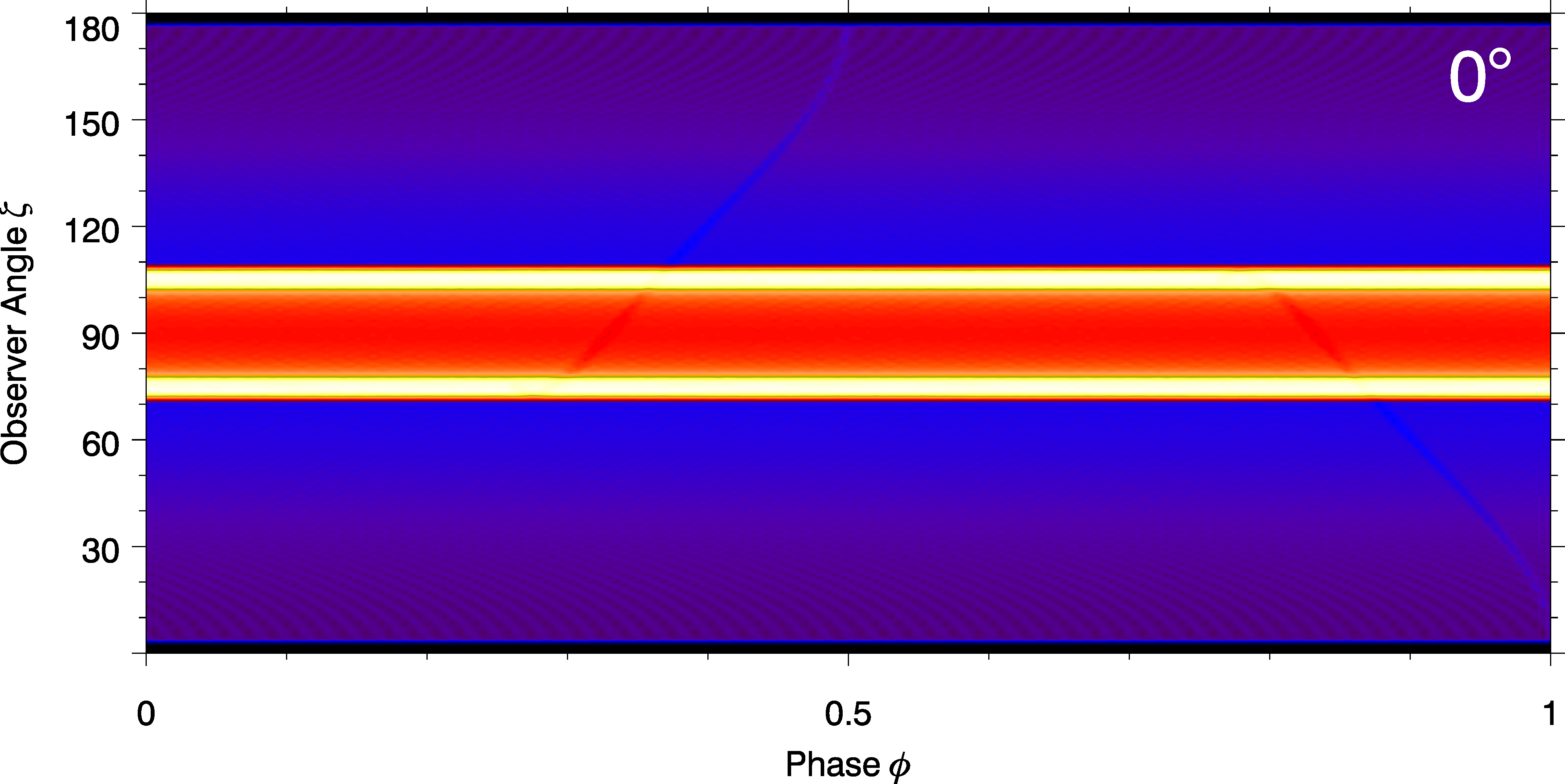}
      \includegraphics[width=\textwidth]{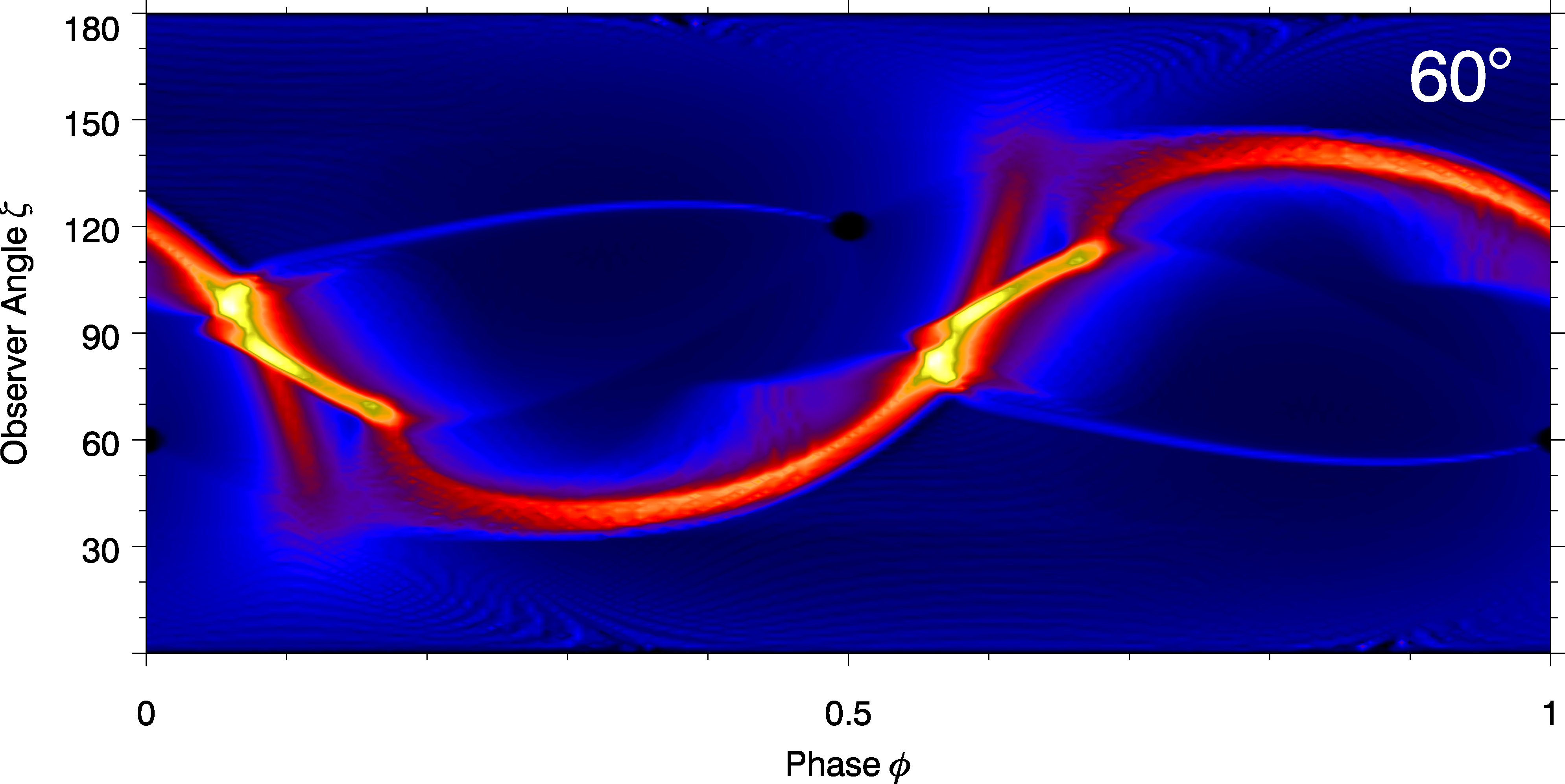}
    \end{minipage}
    \begin{minipage}[r]{0.32\textwidth}
      \includegraphics[width=\textwidth]{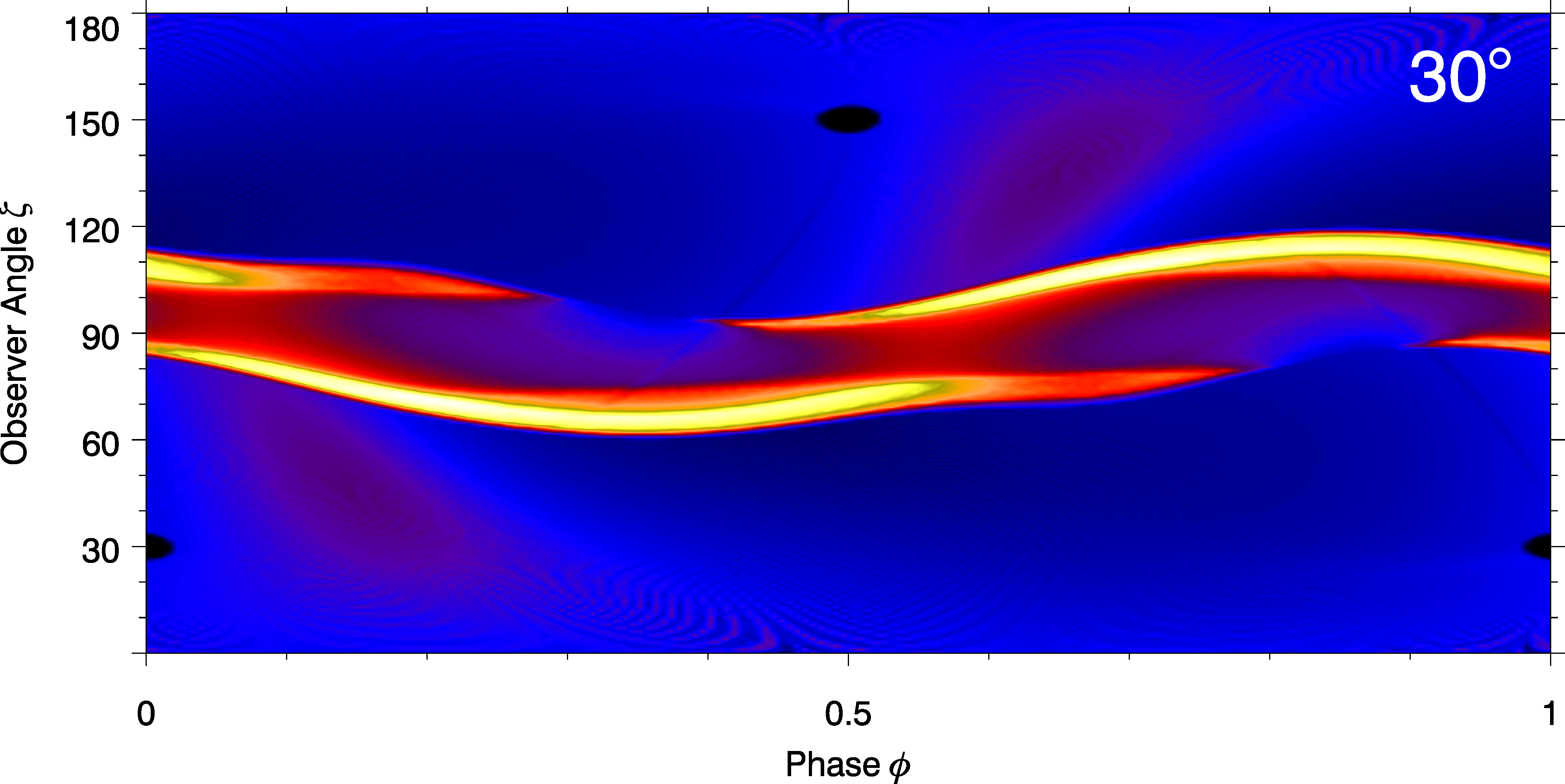}
      \includegraphics[width=\textwidth]{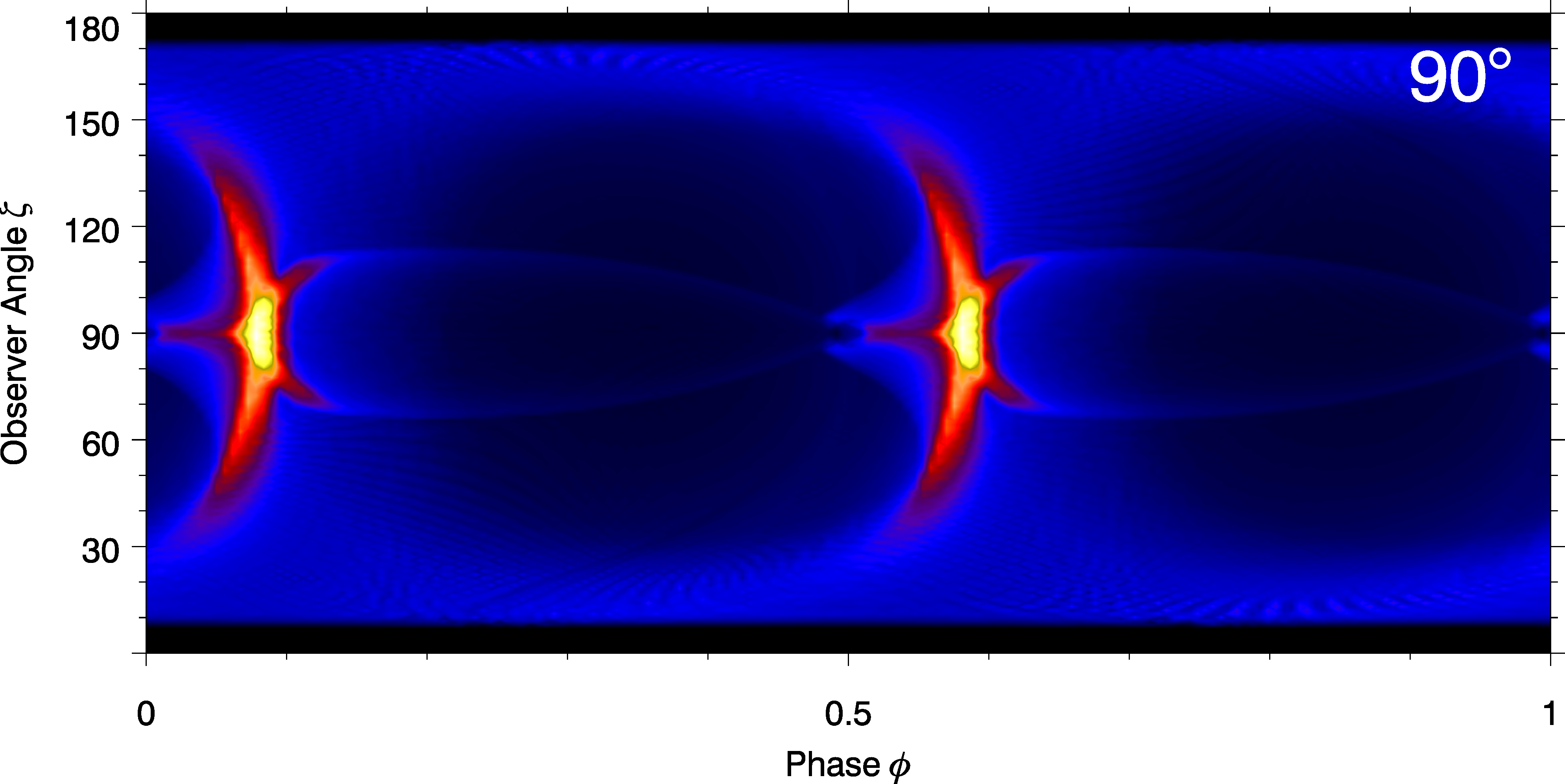}
    \end{minipage}
    \caption{Phaseplots for \al between 0\dg and 90\dg with a resolution of 30\dg in both \al and \z for the TPC model.\label{fig:phaseplots}}
  \end{figure}

    Looking at \figr{phaseplots} we see that the intuition we have built up enables us to very easily pin down some major attributes of the structure of the atlas which would be produced under a certain set of model parameters, in this case a pulsar with a period of 160\,ms, a TPC gap width of 5\%, and a radial gap extent of 1.2\,\rlc.

    We start with a flat and relatively featureless phaseplot at $\al=0\dg$ with the structure of the magnetic field being the only factor determining the structure of the phaseplot. At $\al=30\dg$ we see the secondary caustics and caustics starting to form, as well as the polar cap. We expect to see very few two-peaked profiles, and predominantly profiles with a single early peak associated with the secondary caustic. The caustics are much more prominently visible at $\al=60\dg$, and we can see the notch line and associated overlap region making their appearance. The caustics are more extremely curved, and the secondary caustic and caustics have started to merge. Here we expect to see a much larger number of two-peaked profiles, as well as some very complicated profiles between $\al=50\dg$ and 80\dg, having multiple additional small peaks due to the cut intersecting many of the minor features on the phaseplot. At $\al=90\dg$ the caustics dominate the atlas, yielding a very large range of two-peaked profiles with very large seperations between the peaks ($\bigtriangleup\sim0.5$ in phase). Here the high altitude emissions produce a significant wing (around $(\phi,\z)=(0.5,90\dg)$) leading the main caustic at higher \z, producing LCs with broader peaks.

\section{Discussion and Conclusions}
  In this paper we have illustrated how an understanding of the structure of the phaseplots generated by geometric pulsar models can translate into an understanding of the structure of the resulting LC atlas. Moreover, this understanding also allows us to more easily and effectively understand how changes in the model parameters can effect the LC atlas by understanding how they effect the phaseplots. This approach also allows us to effectively interpret obtained LC fits upon inspection in terms of the features contributing to the best-fit LC, and thus make more successful use of the obtained LC fit in, e.g., predicting at which altitudes the relevant radiation originates.


\ack This research is based upon work supported by the South African National Research Foundation. A.K.H. acknowledges support from the NASA Astrophysics Theory Program. C.V., T.J.J., and A.K.H. acknowledge support from the Fermi Guest Investigator Program.

%

\end{document}